\documentclass[aps,twocolumn,showpacs,preprintnumbers,nofootinbib,prl,superscriptaddress,groupedaddress,10pt]{revtex4-1}

\makeatletter
\def\l@subsubsection#1#2{}
\def\l@subsubsubsection#1#2{}
\makeatother

\usepackage{graphicx,amssymb,amsmath,amsthm,amsfonts,epsfig,epsf}
\usepackage[usenames]{color}
\usepackage{epstopdf}
\definecolor{darkred}{rgb}{0.5,0,0}
\usepackage{aas_macros}
\usepackage{bm}
\usepackage{dcolumn}
\usepackage[latin1]{inputenc}
\usepackage{latexsym}
\usepackage{rotating}
\usepackage{longtable}

\usepackage{enumerate}
\usepackage{tensor,multirow}
\usepackage{url}
\usepackage[linktocpage]{hyperref}

\def\nn{\nonumber}

\def\be{\begin{equation}}
\def\ee{\end{equation}}
\newcommand{\beq}{\begin{eqnarray}}
\newcommand{\eeq}{\end{eqnarray}}

\def\ba{\begin{align}}
\def\ea{\end{align}}
\newcommand{\tn}{\textnormal}

\begin{document}
\title{Probing Planckian corrections at the horizon scale with LISA binaries}

\author{
Andrea Maselli$^{1,2}$,
Paolo Pani$^{3,4}$,
Vitor Cardoso$^{2,5}$, 
Tiziano Abdelsalhin$^{3,4}$,
Leonardo Gualtieri$^{3,4}$,
Valeria Ferrari$^{3,4}$ 
}
\affiliation{$^{1}$ Theoretical Astrophysics, Eberhard Karls University of Tuebingen, Tuebingen 72076, Germany}
\affiliation{${^2}$ Centro de Astrof\'isica e Gravita\c{c}\~ao - CENTRA, Departamento de F\'isica, Instituto Superior T\'ecnico - IST, Universidade de Lisboa - UL, Av. Rovisco Pais 1, 1049-001 Lisboa, Portugal}
\affiliation{$^{3}$ Dipartimento di Fisica, ``Sapienza'' Universit\`a di Roma, Piazzale Aldo Moro 5, 00185, Roma, Italy}
\affiliation{$^4$ Sezione INFN Roma1, Piazzale Aldo Moro 5, 00185, Roma, Italy}
\affiliation{${^5}$ Perimeter Institute for Theoretical Physics, 31 Caroline Street North Waterloo, Ontario N2L 2Y5, Canada}
%
%
\begin{abstract}
Several quantum-gravity models of compact objects predict microscopic or even Planckian corrections at the horizon scale. We explore the possibility of measuring two model-independent, smoking-gun effects of these corrections in the gravitational waveform of a compact binary, namely the absence of tidal heating and the presence of tidal deformability. For events detectable by the future space-based interferometer LISA, we show that the effect of tidal heating dominates and allows one to constrain putative corrections down to the Planck scale. 
The measurement of the tidal Love numbers with LISA is more challenging but, in optimistic scenarios, it allows to constrain the compactness of a supermassive exotic compact object down to the Planck scale. Our analysis suggests that highly-spinning, supermassive binaries at $1-20\,{\rm Gpc}$ provide unparalleled tests of quantum-gravity effects at the horizon scale.
\end{abstract}

\maketitle
%

\noindent{\bf{\em Introduction.}}
%
Gravitational waves (GWs) are the most direct probes of compact objects down to the horizon scale and
can shed light on one of the outstanding open issues in gravitational astronomy:
the nature of compact, dark and massive objects~\cite{Cardoso:2017cqb}. It has been tacitly assumed that the latter must be black holes (BHs) for a number of compelling reasons:
BHs form from classical gravitational collapse of stars, while there are no known, equally-well motivated alternatives which are sufficiently compact
to explain observations, especially the recent GW detections~\cite{Abbott:2016blz,Abbott:2016nmj,Abbott:2017vtc}. Nonetheless, over the years several arguments have been put forward, suggesting that new physics at the horizon scale might set in during gravitational collapse, 
possibly halting or altering the formation of BHs~\cite{Mazur:2004fk,
Mathur:2005zp,Skenderis:2008qn,Gimon:2007ur,Almheiri:2012rt}.
While the end product of the collapse in these scenarios is essentially unknown or model dependent~\cite{Holdom:2016nek,Brustein:2017kcj,Barcelo:2017lnx}, the exotic compact objects (ECOs) that might form share two common features: they are extremely compact and horizonless. Regardless of the viability of these objects and the mechanisms behind them, 
we have, for the first time, the means for testing these scenarios with GWs.

It was recently argued that ECOs can be detected or ruled out through GW measurements in two different regimes: the postmerger ringdown phase of a coalescence --~where putative corrections at the horizon scale will produce GW echoes~\cite{Cardoso:2016rao,Cardoso:2016oxy,Maselli:2017je} (see also Ref.~\cite{Ferrari:2000sr} for an earlier study, and Refs.~\cite{Abedi:2016hgu,Ashton:2016xff,Abedi:2017isz} for a debate on the evidence of this effect in aLIGO data)~-- and the late-time inspiral of the coalescence --~through the measurement of the tidal deformability of the two objects~\cite{Wade:2013hoa,Cardoso:2017cfl,Sennett:2017etc} and of their spin-induced quadrupole moment~\cite{Krishnendu:2017shb}.

Here, we discuss another effect that can be used to distinguish ECOs from BHs, namely the absence of tidal heating if the binary components do not possess a horizon. For supermassive binaries detectable by the future Laser Interferometer Space Antenna (LISA)~\cite{LISA}, we show that this effect can be used to constrain putative corrections near the horizon down to the Planck scale, even for sources at cosmological distance\footnote{We assume a standard $\Lambda$CDM flat universe with $H_0\sim67.7$, $\Omega_M\sim0.26$ and $\Omega_\Lambda\sim0.69$~\cite{Ade:2015xua}. All masses quoted in the text are assumed to be the redshifted masses. With this choice, the waveform~\eqref{waveform} is independent of the redshift $z$.} if the binary components are highly spinning, as predicted by several models of spin evolution of supermassive objects~\cite{Sesana:2014bea}.

\noindent{\bf{\em GW tests of the nature of compact objects.}}
%
Consider a compact binary, of masses $m_i$ ($i=1,2$), total mass $m=m_1+m_2$, mass ratio $q=m_1/m_2\geq1$, and dimensionless spins $\chi_i$.
In a post-Newtonian (PN) approximation (i.e. a weak-field/slow-velocity expansion of Einstein's equations), dynamics is driven by energy and angular momentum loss, and particles are endowed with a series of multipole moments and with finite-size tidal corrections~\cite{Blanchet:2006zz}. 
Loosely speaking, the nature of the inspiralling objects is encoded in (i) the way they respond to their own field --~i.e., on their own multipolar structure, (ii) the way they respond when acted upon by the external gravitational field of their companion~-- through their tidal Love numbers (TLNs)~\cite{PoissonWill}, and (iii) on the amount of radiation that they possibly absorb, i.e. on tidal heating~\cite{Hartle:1973zz,PhysRevD.64.064004}.
These effects are all included in the waveform produced during the inspiral, and can be incorporated in the (Fourier-transformed) GW signal as (we use $G=c=1$ units)
\begin{equation}
\tilde{h}(f)={\cal A}(f)e^{i(\psi_\tn{PP}+\psi_\tn{TH}+\psi_\tn{TD})}\ , \label{waveform}
\end{equation}
where $f$ and ${\cal A}(f)$ are the GW frequency and amplitude, $\psi_\tn{PP}(f)$ is the ``pointlike'' phase, whereas $\psi_\tn{TH}(f),\,\psi_\tn{TD}(f)$ are the contribution of the tidal heating and the tidal deformability, respectively.

Spin-orbit and spin-spin interactions are included in $\psi_\tn{PP}$, the latter also depending on the spin-induced quadrupole moment. This property has been recently used to constrain ${\cal O}(\chi_i^2)$ deviations from the Kerr geometry~\cite{Krishnendu:2017shb}. However, in known models of rotating ECOs --~e.g., gravastars~\cite{Pani:2015tga,Uchikata:2016qku} and strongly-anisotropic, incompressible neutron stars~\cite{Yagi:2015hda}~-- the multipole moments approach those of a Kerr BH in the high-compactness limit. 
This suggests that the distinction between ultracompact objects and BHs can only be done, realistically, using finite-size corrections, $\psi_\tn{TH}(f)$ and $\psi_\tn{TD}(f)$.

BHs are very special objects in general relativity and it is no surprise that finite-size effects are different for ECOs than for BHs. 
If the inspiralling objects are BHs, a small contribution to the dynamics is provided by dissipation of energy and angular momentum at the horizon. To leading PN order, the energy flux at the horizon reads~\cite{Alvi:2001mx,PhysRevD.64.064004,Taylor:2008xy,Poisson:2009di,Nagar:2011aa,Bernuzzi:2012ku,Chatziioannou:2012gq,Cardoso:2012zn} (henceforth we assume circular orbits and spins aligned with the orbital angular momentum)
\begin{eqnarray}
 \dot{E}_{\rm BH}^{\rm heating} &=&-\dot{E}_{\rm GW}\sum_{i=1,2} \frac{v^5}{4}\left(\frac{m_i}{m}\right)^3(1+3\chi_i^2)\nn\\
 &\times&\left\{\chi_i-2\left[1+\sqrt{1-\chi_i^2}\right]\frac{m_i}{m}v^3\right\} \,, \label{Edot}
\end{eqnarray}
where the expansion parameter $v=(\pi m f)^{1/3}$ is the orbital velocity and $\dot{E}_{\rm GW}\sim v^{10}$ is the leading order, quadrupole GW flux~\cite{Peters:1963ux}. Angular-momentum flux is subleading
so the spins remain roughly constant during the evolution~\cite{Alvi:2001mx}.
The GW phase $\psi$ is governed by $d^2\psi/df^2 =2\pi(dE/df)/\dot E$, where $E\sim v^2$ is the binding energy of the binary. To
the leading order, this yields
\begin{equation}\label{Hphase}
\psi_\tn{TH}^{\rm BH} = \psi_{\tn{N}}\left(F(\chi_i,q) v^5\log v+G(q) v^8 [1-3\log v]\right)\,,
\end{equation}
where $\psi_{\tn{N}}\sim v^{-5}$
is the leading-order contribution to the point-particle phase (corresponding to the flux ${\dot
  E}_{\rm GW}$), $F(\chi_i,q)$ and $G(q)$ are simple functions of their arguments, and $F\sim\chi_i+{\cal O}(\chi_i^2)$.
Thus, absorption at the horizon introduces a $2.5$PN ($4$PN)$\times\log v$ correction to the GW phase of spinning (nonspinning)
binaries, relative to the leading term (conventionally, one PN order corresponds to a term $\sim v^2$). For all known matter, GW absorption is negligible: tidal heating is therefore a good discriminator
for the existence of horizons.
It is therefore convenient to introduce an absorption coefficient $\gamma$ such that, $\psi_\tn{TH}\equiv \gamma\, \psi_\tn{TH}^{\rm BH} $, with $\gamma=0$ for ECOs with a perfectly reflecting surface or whose interior does not absorb GWs, $\gamma=1$ for BHs, and $\gamma\in(0,1)$ for partial absorption.
\begin{figure}[th]
\centering
\includegraphics[width=0.41\textwidth]{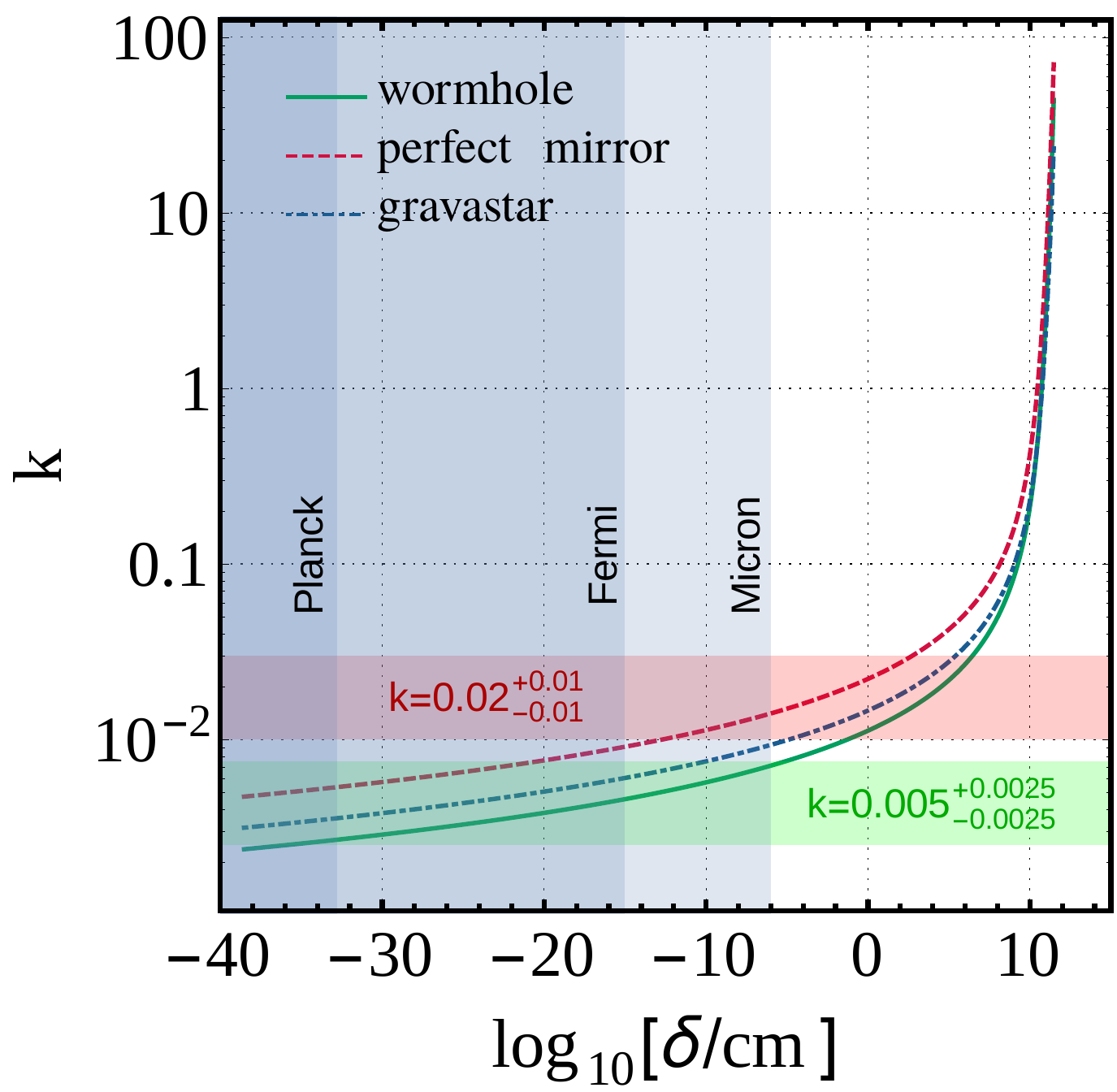}
\caption{TLN of ECOs~\cite{Cardoso:2017cfl} as a function of the distance $\delta=r_0-2M$ between the surface and the Schwarzschild radius, for an ECO mass $M=10^6 M_\odot$. From left to right, vertical strips mark different scales: Planck, subnuclear (Fermi), and microscopic scales, respectively. The horizontal bands correspond to the measurement of TLNs at the level of $k\approx0.02$ (as routinely achievable by LISA, cf.\ Fig.~\ref{fig:bounds} below) and $k\approx0.005$ (achievable only in optimistic scenarios).}
\label{fig:Love}
\end{figure}

In addition, while the TLNs of BHs are zero~\cite{Binnington:2009bb,Damour:2009vw,Fang:2005qq,Gurlebeck:2015xpa,Poisson:2014gka,Pani:2015hfa}, those of ECOs are small but finite~\cite{Pani:2015tga,Uchikata:2016qku,Porto:2016zng,Cardoso:2017cfl}. In line with neutron star binaries~\cite{PoissonWill,Flanagan:2007ix,Hinderer:2007mb}, 
the leading tidal deformability term for ECO binaries reads,
\begin{equation}
\psi_\tn{TD}(f)=- \psi_{\tn{N}}\frac{\Lambda}{6m^{5}} v^{10}\frac{(1+q)^2}{q}\,,
\end{equation}
where $\Lambda=\left(1+{12}/{q}\right)m_1^5 k_1+(1+12 q)m_2^5 k_2$ is the weighted tidal deformability, and $k_i$ is the (dimensionless) TLN of the $i-$th object.
Thus, tidal deformability introduces a $5$PN correction to the GW phase relative to the leading-order GW term, whereas spin-tidal couplings are subleading~\cite{Poisson:2014gka,Pani:2015nua,Landry:2015zfa} and can be neglected.

The TLNs of a nonspinning ultracompact object of mass $M$ and radius $r_0=2M(1+\epsilon)$ (with $\epsilon\ll1$) in Schwarzschild coordinates vanish logarithmically in the BH limit~\cite{Cardoso:2017cfl}, $k\sim 1/|\log\epsilon|$, opening the way to probe horizon scales. As depicted in Fig.~\ref{fig:Love}, any measurement of the TLN translates into an estimate of the distance of the ECO surface from its Schwarzschild radius,
\begin{equation}
 \delta:=r_0-2M\sim 2Me^{-1/k}\,.
\end{equation}
For a supermassive object with $M\sim 10^6 M_\odot$, $\delta$ is of the order of the Planck length, $\ell_P\approx 1.6\times 10^{-33}\,{\rm cm}$, when $k\approx 0.005$ [cf.\ Fig.~\ref{fig:Love}]. Therefore, future GW observations should aim at reaching the level of accuracy necessary to measure TLNs as small as $k\sim0.005$. Below, we show that this will be achievable with LISA.

To summarize, finite-size effects in the inspiral waveform provide two different null-hypothesis tests of BHs. While BHs have zero TLNs but introduce a nonzero tidal heating ($\psi_\tn{TD}=0$, $\gamma=1$), ECOs have (logarithmically small) TLNs but zero tidal heating ($\psi_\tn{TD}\neq0$, $\gamma=0$).

\begin{figure*}[th]
\centering
\includegraphics[width=0.49\textwidth]{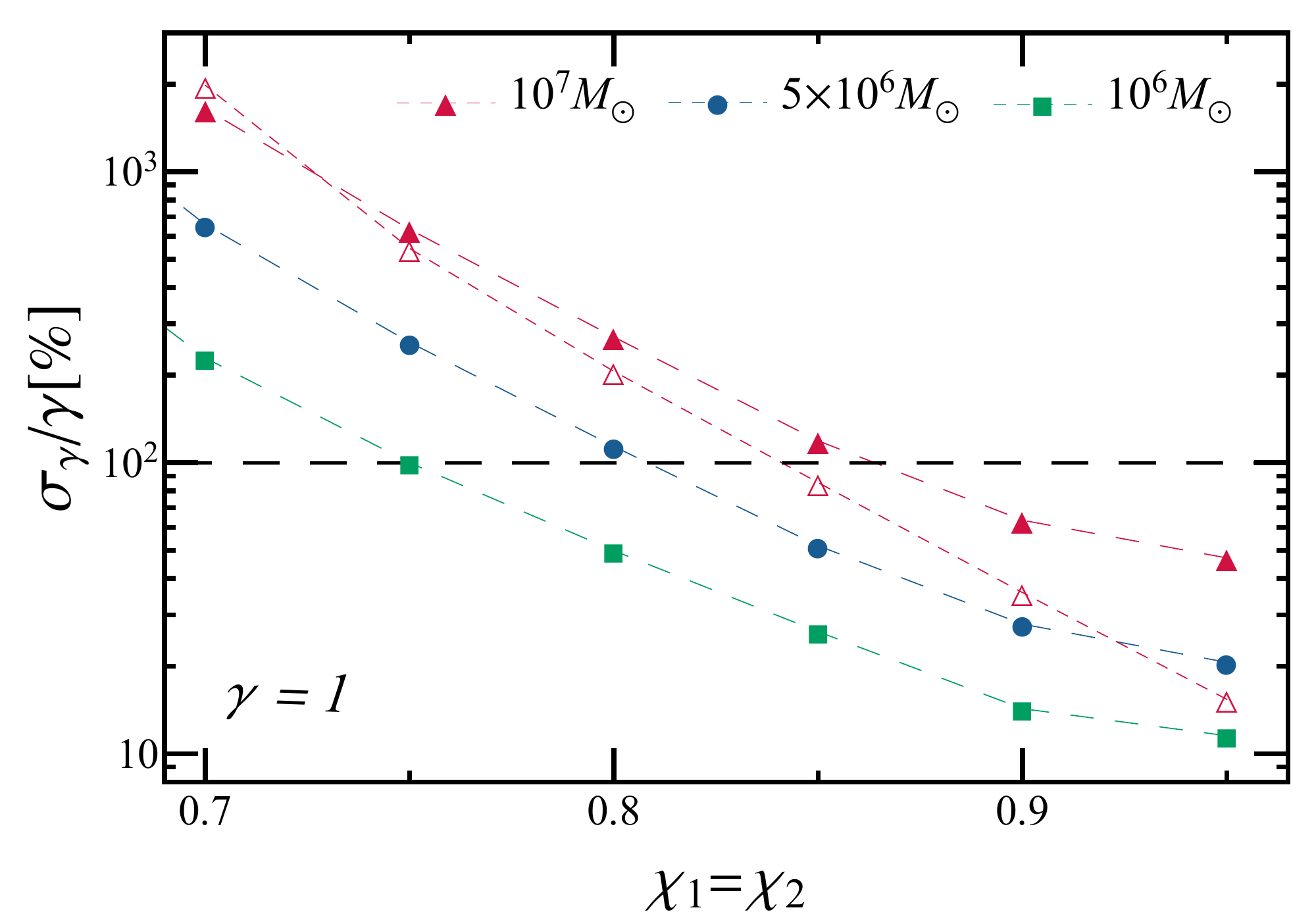}
\includegraphics[width=0.49\textwidth]{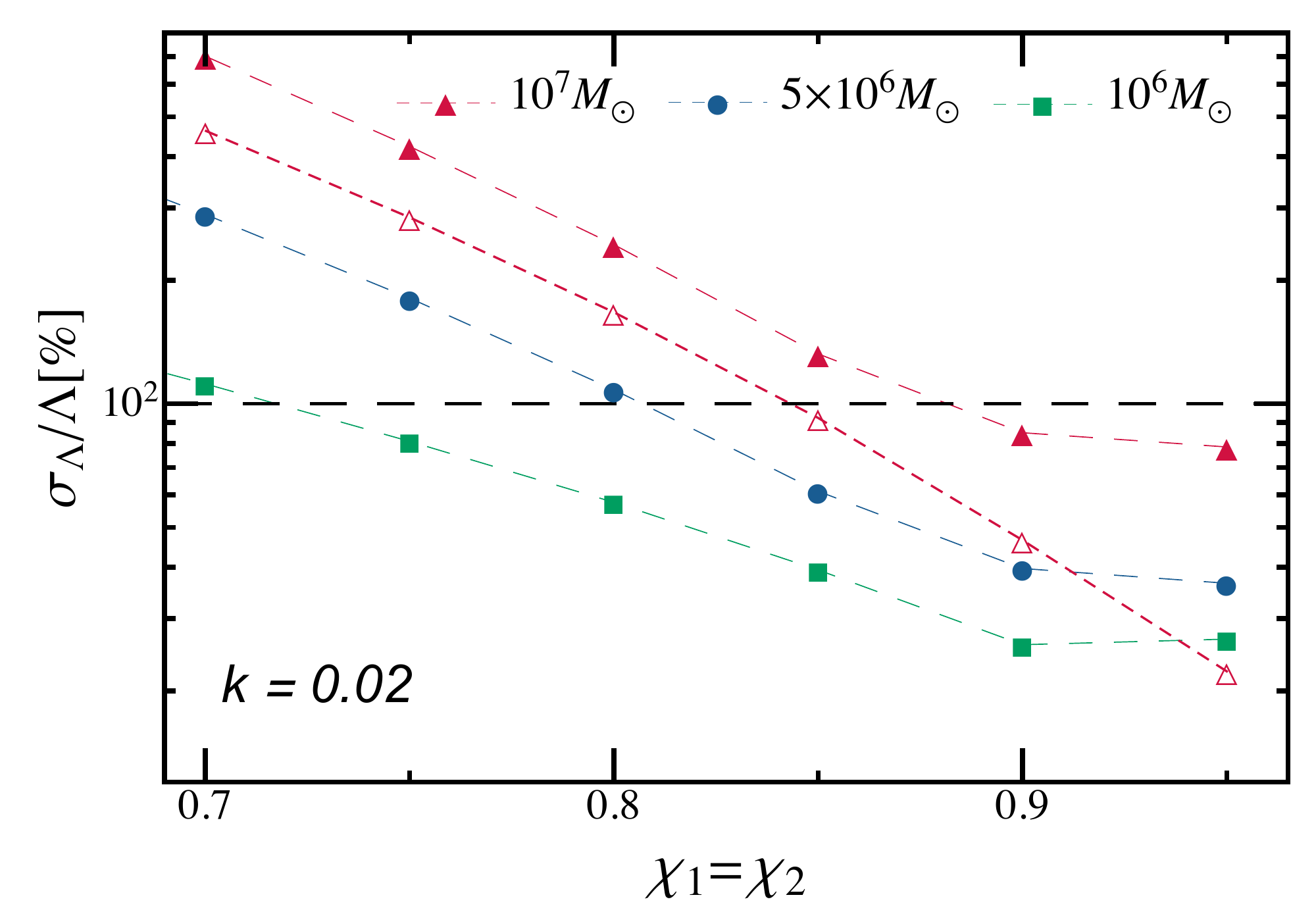}
\caption{Percentage relative errors on the tidal-heating parameter $\gamma$ (left panel) and on the average tidal deformability $\Lambda$ (right panel) as a function of the spin parameter $\chi_{1}=\chi_2$, 
for different values of the mass $m_1=(10^6,5\times 10^6,10^7)M_\odot$.
In the left panel we consider a BH binary, i.e.\ we assume $\gamma=1$ and $\Lambda=0$, whereas in the right panel we assume an ECO binary (i.e., $\gamma=0$ and $\Lambda\neq0$, with $k=0.02$ for both binary components). 
Full (empty) markers refer to mass ratio $q=1.1$ ($q=2$).
Points below the horizontal line correspond to detections that can distinguish between a BH and an ECO at better than $1\sigma$ level.
We assume binaries at luminosity 
distance $2\,{\rm Gpc}$; $\sigma_{\Lambda,\gamma}$ scales with the inverse luminosity distance, whereas $\sigma_\Lambda$ ($\sigma_\gamma$) scales with $1/\Lambda$ ($1/\gamma$) when $k\ll1$ ($\gamma\ll1$). 
}
\label{fig:bounds}
\end{figure*}
%

\noindent{\bf{\em Error analysis.}}
%
State-of-the-art waveform models for BH binaries are fitted to numerical simulations in order to reproduce precisely the late-inspiral and merger signal~\cite{Taracchini:2013rva,Husa:2015iqa,Khan:2015jqa}. Since simulations of ECO binaries are not yet available [with the notable exception of boson stars~\cite{Bezares:2017mzk}, which are however limited to $\epsilon\gtrsim{\cal O}(1)$], there is no fully consistent inspiral-merger-ringdown waveform model for an ECO coalescence.

To overcome this problem, here we focus our analysis on the inspiral phase only, adopting
a state-of-the-art PN template in the frequency domain, the so-called TaylorF2 approximant~\cite{Damour:2000zb,Damour:2002kr,Arun:2004hn}.
We include spin-orbit, spin-spin~\cite{Khan:2015jqa} and the known cubic spin corrections to $\psi_{\rm PP}$ up to $3.5$PN order~\cite{Isoyama:2017tbp}.
As discussed, we assume the quadrupole moment to be that of a Kerr BH.
We include tidal heating to leading order (i.e., to $2.5$PN ($4$PN) relative order for spinning (nonspinning) binaries) and tidal deformability terms to next-to-leading order (i.e, to $6$PN relative order~\cite{PhysRevD.83.084051,PhysRevD.85.124034}).

We employ a Fisher matrix analysis, which is accurate at large signal-to-noise ratios (SNRs)~\cite{Vallisneri:2007ev}, as those relevant for LISA supermassive binaries under consideration.
For a given set of parameters $\vec{\xi}$, the error associated with the measurement of parameter $\xi^a$ is $\sigma_{a}=\sqrt{\Sigma^{aa}}$, where $\Sigma^{ab}$ is the inverse of the Fisher matrix, $\Gamma_{ab}=\left(\partial_{\xi^a} h \vert \partial_{\xi^b} h\right)_{\vec{\xi}=\vec{\xi}_0}$, $\vec{\xi}_0$ are the injected values of the parameters $\vec{\xi}$, and the inner product is defined as
\begin{equation}\label{inner}
(g\vert h)=4\,{\rm Re}\int_{f_{\rm min}}^{f_{\rm max}}df\,\frac{\tilde h(f)\tilde g^{\star}(f)}{S_{h}(f)}\ ,
\end{equation}
where $S_h(f)$ is the recently proposed LISA's noise spectral density~\cite{LISA}. We assume an observation time $T_\tn{obs}=1\,\tn{yr}$, sky-averaging the GW signal~\cite{Berti:2004bd}, and estimate the initial frequency $f_{\rm min}$ of the binary in terms of $f_\tn{max}$ and $T_{\rm obs}$ by solving the binary motion to leading order.
Since higher-PN corrections are more relevant in the last stages of the inspiral, we are interested in those ``golden'' binaries which remain in the LISA band for $T_{\rm obs}$ up until the merger.
Finally, we choose $f_{\rm max}$ to guarantee that our template is a reliable approximation of the signal in the relevant frequency range. Namely, we choose $f_{\rm max}$ such that the overlap ${\cal O}$ between the TaylorF2 BH-BH template and a more accurate inspiral-merger-ringdown template for spinning BH binaries (the PhenomD waveform~\cite{Husa:2015iqa,Khan:2015jqa}) is at least equal to the fiducial threshold ${\cal O}\geq0.95$ (more details are given in the Supplement Material).
We note that, to the leading order, the statistical errors estimated through the Fisher matrix are independent of the
systematic errors arising from approximating the true signal with an imperfect theoretical template~\cite{Chatziioannou:2014bma}.

The relevant Fisher-matrix parameters are $\vec{\xi}=(\ln{\cal A},\psi_c,t_c,\ln {\cal M},\ln\nu,\chi_1,\chi_2)$ [where $\psi_c$, $t_c$, ${\cal M}=(m_1m_2)^{3/5}/m^{1/5}$ and $\nu=m_1m_2/m^2$ are the phase and time at the coalescence, the chirp mass, and the symmetric mass ratio, respectively] plus possibly $\gamma$ and $\log\Lambda$, depending on the system under consideration. (The subleading term of $\psi_{\rm TD}$ depends on an extra parameter~\cite{PhysRevD.85.124034} which, however, turns out to be unmeasurable.)

\noindent{\bf{\em GW constraints on ECOs.}}
Our requirement ${\cal O}\geq0.95$ is not satisfied for \emph{both} large spins and large mass ratios. Because the effects of the former are more important, we focus on $q\in(1,2)$ and $\chi_i\in(0,0.95)$. 

Let us first consider the effects of tidal heating, by setting $\Lambda=0$. 
The error $\sigma_\gamma$ is shown in the left panel of Fig.~\ref{fig:bounds} as a function of the spin for different systems. The dashed horizontal line marks the threshold $\sigma_\gamma=1$: measurements of the tidal heating parameter $\gamma$ below the threshold have less than $100\%$ uncertainty and would discriminate between ECOs~($\gamma=0$) and BHs~($\gamma=1$) at least at $1\sigma$ level. 
While slowly-spinning, equal-mass binaries are not distinguishable, the accuracy improves dramatically at large spin. 
In some favorable scenarios, BH coalescences up to luminosity distances of $20\,{\rm Gpc}$ (corresponding to redshift $z\approx2.5$) with individual masses $m_i\sim10^6 M_\odot$ and spins $\chi_i\gtrsim0.9$ can be confidently distinguished from ECO binaries on the basis of the presence of tidal heating.  
The enhancement with spin is expected, given that tidal heating enters the GW phase at $2.5$PN$\times\log v$ order [cf.\ Eq.~\eqref{Hphase}] and that the merger frequency is higher. We also observe a milder improvement as $q$ increases, due to the larger number of orbits in bandwidth.
One of our best-case scenarios ($m_1=1.1 m_2= 10^6M_\odot$, $\chi_i=0.9$) corresponds to ${\rm SNR}\approx 2\times10^4$, and 
relative errors $\simeq(2\times10^{-3},6\times 10^{-2},6,7,14)\%$ on $({\cal M},\nu,\chi_1,\chi_2,\gamma)$, respectively.

The second case of interest is orthogonal to the above: with ECO binaries in mind we now set $\gamma=0=\psi_{\rm TH}$ but a nonvanishing TLN. We are interested in estimating whether a measurement of $\Lambda$ is incompatible with zero and therefore whether ECOs can be distinguished from BHs on these grounds.
This case was preliminarily studied in Ref.~\cite{Cardoso:2017cfl} (and more recently in Ref.~\cite{Sennett:2017etc}) only for equal-mass binaries and neglecting spin. The right panel of Fig.~\ref{fig:bounds} shows that the constraints on $\Lambda$ are orders of magnitude more stringent for spinning binaries. In particular, for $\chi_i\gtrsim0.9$ and $q\sim1$, a constraint on the TLN as stringent as $k\lesssim 0.005$ ($k\lesssim 0.05$) can be obtained for a coalescence within $2\,{\rm Gpc}$ ($20\,{\rm Gpc}$).
For several ECO models~\cite{Cardoso:2017cfl}, a bound $k\lesssim0.005$ on the TLN translates into an impressive constraint on the compactness of ECOs down to the Planck scale near the horizon, i.e. $\delta/M\sim 10^{-45}$ for a supermassive object [cf.\ Fig.~\ref{fig:Love}].

Thus, finite-size effects open up the tantalizing possibility to know, through GW observations, if BHs do actually exist, and are in fact complementary to searches for new physics at the horizon scale through the detection of GW echoes~\cite{Cardoso:2017njb}.

\noindent{\bf{\em Effective absorption by ECOs.}}
%
It might be argued that there should be a continuous transition from BHs to ultracompact horizonless objects, whereas we have assumed that tidal heating of ECOs is negligible. In fact, an ultracompact object can trap radiation within its photon sphere efficiently~\cite{Cardoso:2014sna,Cardoso:2016rao,Cardoso:2016oxy}, thus mimicking the effect of a horizon. This mechanism is interesting \emph{per se} at a theoretical level and deserves (and requires) a separated study. Nonetheless, we can estimate it as follows. In order for the absorption to affect the orbital motion, it is necessary that the (arrival) time radiation takes to reach the companion, $T_{\rm arr}$, be much longer than the radiation-reaction time scale due to heating, $T_{\rm RR}\sim E/\dot E_{\rm BH}^{\rm heating}$. From Eq.~\eqref{Edot}, the latter reads at most $T_{\rm RR}\sim m v^{-13}/(\chi_i+3\chi_i^3)$, when $q=1$ and $\chi_i\neq0$.

For BHs, $T_{\rm arr}\to\infty$ because of time dilation, so that the condition $T_{\rm arr}\gg T_{\rm RR}$ is always satisfied. 
For ECOs in the $\epsilon\to0$ limit, $T_{\rm arr}$ turns out to be just the GW echo delay time of the $i$-th object, $T_{\rm arr}\sim m_i|\log{\epsilon}|$~\cite{Cardoso:2016rao,Cardoso:2016oxy}. For a given compactness, the condition $T_{\rm arr}\gg T_{\rm RR}$ implies that an effective heating can take place only when $f\gg f_{\rm crit}$. Including spins~\cite{Abedi:2016hgu} and generic mass ratios in the computation, we find,
\begin{equation}
 f_{\rm crit}\sim \frac{2^{1/13}}{2 \pi m} \left[\frac{5mQ\Delta_i}{m_i \left(\chi _i+3 \chi
   _i^3\right) \left(1+\Delta_i\right) \log(r_{+}^{(i)}/\delta)}\right]^{3/13}\,, 
\end{equation}
where $\Delta_i =\sqrt{1-\chi _i^2}$, $Q=(1+q)^4/[q (1+q(q-1))]$, and $r_+^{(i)}=m_i(1+\Delta_i)$. When $\chi_i\to0$, the subleading term in Eq.~\eqref{Edot} becomes dominant, but its effect is negligible.
As expected, $f_{\rm crit}$ decreases as the compactness increases. However, owing to the logarithmic dependence, even for $\delta\sim \ell_P$,  $f_{\rm crit}\gtrsim f_{\rm ISCO}$ when $\chi_i<0.9$ or when $q\gg1$ for any spin. In the least favorable case [$0.9<\chi_i<0.99$ and $q\sim1$], $f_{\rm crit}\approx 0.4f_{\rm ISCO}$. 
Therefore, in the entire region where the PN expansion is valid the ``effective'' tidal heating of a Planck-scale ECO can be neglected.

Another possible caveat concerns GW dephasing due to mode excitation. The fundamental QNMs of an ECO have small frequency and are extremely long lived. Because $\omega_{\rm QNM}\sim {2\pi}/{T_{\rm arr}}\sim 1/(M|{\log\epsilon}|)$~\cite{Cardoso:2016rao,Cardoso:2016oxy}, the QNMs can be possibly excited only when the orbital frequency $\simeq 2\pi/T_{\rm arr}$ which, for Planck-scale ECOs, occurs only near the ISCO and with extremely narrow resonances that should absorb a negligible amount of energy. 

In summary, our results seem robust even when relaxing some of the assumptions:
mode excitation and tidal heating can be expected to be negligible for Planck-scale ECOs, validating our analysis.

\noindent{\bf{\em Discussion and Prospects.}}
%
The future space-based GW detector LISA~\cite{LISA} will be able to distinguish whether supermassive dark objects in a binary coalescence have a horizon or not by measuring two distinct and complementary finite-size effects on the waveform: tidal heating and tidal deformability. These effects become stronger for highly spinning binaries --~as those predicted in several models of BH spin cosmic evolution~\cite{Sesana:2014bea}~-- and allow us to constrain the location of the ECO surface down to Planck scales, even for cosmological sources. This is a truly spectacular potential. GW observations will possibly provide the most impressive tests of near-horizon physics, and will challenge our understanding of quantum gravity backreaction effects.

This work is intended as a proof-of-principle; including eccentricity, spin misalignment, higher-order PN terms,
and performing a Bayesian analysis are certainly relevant extensions. 
In the near future, we expect that a substantial effort will be devoted to perform fully relativistic simulations of exotic binaries. The latter would be extremely 
precious to extend the domain of validity of current waveforms to the merger and ringdown phases, matching semi-analytical templates to the numerical data, in order 
to build complete ECO templates (e.g. using the hybrid waveform approach~\cite{Husa:2015iqa,Khan:2015jqa}, or an effective-one-body
framework~\cite{Buonanno:1998gg,Bernuzzi:2012ku,Taracchini:2013wfa,Taracchini:2013rva}).

LISA rates of these events at $z\lesssim2.5$ are uncertain, but they are expected to be $\sim1-10$ per year~\cite{Klein:2015hvg}. 
Since even a single detection of such systems during LISA lifetime is enough to impose exquisite constraints, we claim that highly spinning, supermassive binaries may be wonderful probes of putative quantum-gravity effects at the horizon scale.

\noindent{\bf{\em Acknowledgments.}}
%
We are grateful to Leor Barack and Eric Poisson for useful correspondence regarding self-force effects in ultracompact backgrounds,
to Ryuichi Fujita for clarifying the convergence properties of the PN series in the extreme-mass ratio limit, to Emanuele Berti for interesting discussions on the Fisher-matrix analysis, and to Michele Vallisneri for useful correspondence and for comparing with us the results of Ref.~\cite{Vallisneri:2007ev}.
We are indebted to Tanja Hinderer and Scott Hughes for a critical reading of the manuscript and for many useful suggestions for improvement.
We acknowledge interesting discussions with the participants of the workshops ``New Frontiers in Gravitational-Wave Astrophysics'' and ``Strong Gravity Universe'', especially with Alessandra Buonanno, Kent Yagi, and Nico Yunes.
P.P.\ acknowledges financial support provided under the European Union's H2020 ERC, Starting Grant agreement no.~DarkGRA--757480 and is grateful to the Mainz Institute for Theoretical Physics (MITP) for its hospitality and its partial support during the completion of this work.
V.C. acknowledges financial support provided under the European Union's H2020 ERC Consolidator Grant ``Matter and strong-field gravity: New frontiers in Einstein's theory'' grant agreement no. MaGRaTh--646597. Research at Perimeter Institute is supported by the Government of Canada through Industry Canada and by the Province of Ontario through the Ministry of Economic Development $\&$
Innovation.
The authors would like to acknowledge networking support by the COST Action CA16104.
This project has received funding from the European Union's Horizon 2020 research and innovation programme under the Marie Sklodowska-Curie grant agreement No 690904, the ``NewCompstar'' COST action MP1304, and from FCT-Portugal through the projects IF/00293/2013.
The authors thankfully acknowledge the computer resources, technical expertise and assistance provided by CENTRA/IST. Computations were performed at the clusters
``Baltasar-Sete-S\'ois'' and Marenostrum, and supported by the MaGRaTh--646597 ERC Consolidator Grant.
%

\appendix

\section{Reliability of the TaylorF2 template}
\begin{figure*}[th]
\centering
\includegraphics[width=0.36\textwidth]{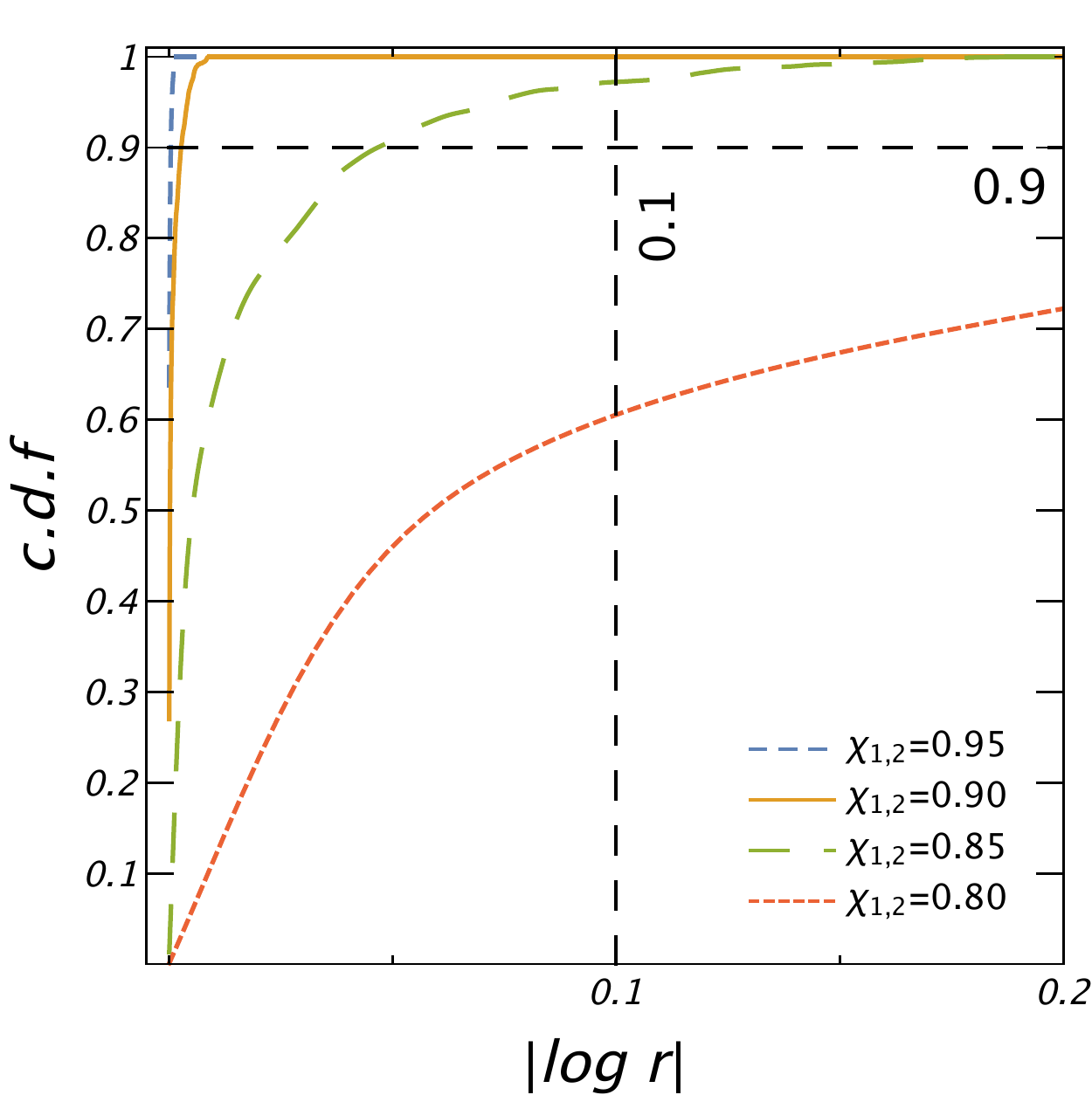}
\includegraphics[width=0.36\textwidth]{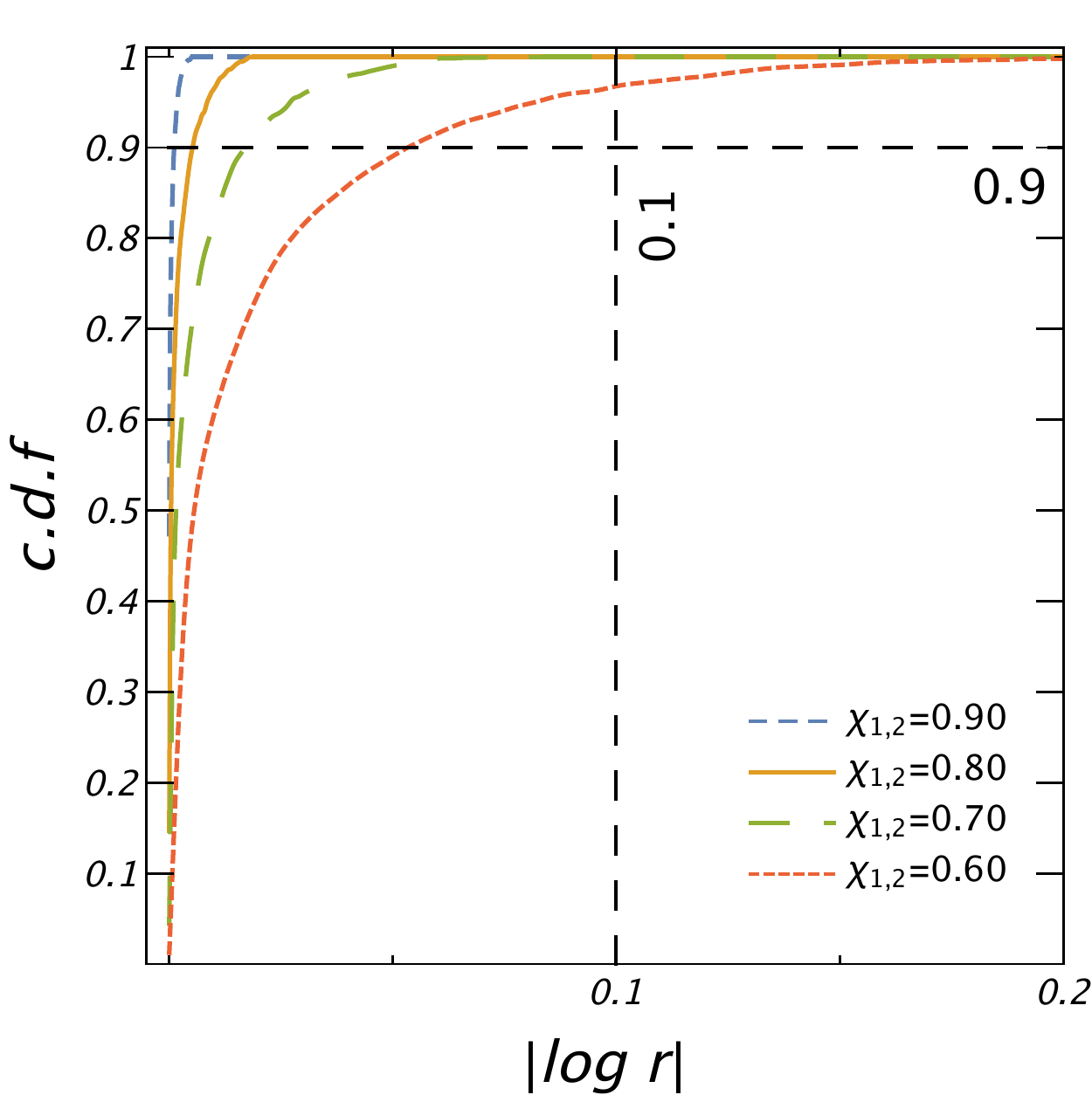}
\caption{Comulative distribution function of the maximum mismatch $\vert \ln r\vert$ as defined in Ref.~\cite{Vallisneri:2007ev} for the most representative cases shown in Fig.~2 for the tidal heating (left panel) and the TLN (right panel) contribution. We considered $m_1=10^6 M_\odot$, $q=1.1$, $d=2\,{\rm Gpc}$, and different values of $\chi_1=\chi_2$.
}
\label{fig:logr}
\end{figure*}

In order to guarantee that the TaylorF2 template adopted in our analysis provides a reliable approximation of the inspiral waveform in the relevant frequency range, we compare it to the PhenomD waveform~\cite{Husa:2015iqa,Khan:2015jqa}; the latter provides an accurate inspiral-merger-ringdown template for spinning BH binaries. We choose $f_{\rm max}$ such that the overlap ${\cal O}\equiv {\rm max}_{t_c,\phi_c}(h_{\rm TF2}|h_{\rm PD})>0.95$, where $h_{\rm TF2}$ is our waveform in the BH-BH case (i.e., $\gamma=1$, $k_i=0$), whereas $h_{\rm PD}$ is the PhenomD template for the same parameters.

For all results presented in this work, the overlap between our model and PhenomD is at least ${\cal O}\gtrsim0.95$ (and often ${\cal O}\gtrsim0.99$). We have also checked that using a more stringent threshold (${\cal O}\gtrsim0.97$)
would increase the statistical errors $\sigma_\gamma$ and $\sigma_\Lambda$ just by a factor of a few in the optimal configurations.
A less conservative choice (often adopted in previous exploratory studies) is to set a sharp cut-off, $f_{\rm max}=\min(1\,{\rm Hz},f_{\rm ISCO})$, where $f_{\rm ISCO}$ is the GW frequency at the innermost-stable circular orbit~(ISCO) of the Kerr metric\footnote{Because the external geometry of an ECO is arbitrarily close to that of a Kerr BH when $\epsilon\to0$, the Kerr ISCO provides a good estimate of the cutoff.}, including corrections due to the self-force and the spin of the less massive object~\cite{Favata:2010ic}. This choice would provide more stringent constraints than those reported in the main text and, to the leading order, would not affect the statistical errors estimated through the Fisher matrix~\cite{Chatziioannou:2014bma}.

Finally, we have performed a consistency check of our numerical approach, computing the maximum mismatch $\vert \ln r\vert$, 
i..e. the ratio between the Fisher-approximate likelihood and the exact probability distribution \cite{Vallisneri:2007ev}. Specifically, the 
mismatch creterion quantifies whether the strength of the signal is large enough to justify the high SNR approximation, i.e. whether 
the Fisher matrix can be consistently used to represent a given parameter's uncertainty. Following \cite{Vallisneri:2007ev}, we have 
computed $\vert \ln r\vert$ requiring that $90\%$ of its cumulative distribution is below the threshold $\vert \ln r\vert<0.1$. 
The result of this analysis is shown in Fig.~\ref{fig:logr}, for some of the most representative configurations below the threshold of Fig.~2. This analysis indicates that the statistical errors estimated from the Fisher matrix should be a reliable estimate of the measurement precision achievable by LISA for the sources under consideration.

\bibliographystyle{apsrev4}
\bibliography{refs}
	
\end{document}